# Disease2Vec: Representing Alzheimer's Progression via Disease Embedding Tree

Lu Zhang[1*], Li Wang[12], Tianming Liu[3], Dajiang Zhu[1]

[1]Computer Science and Engineering, University of Texas at Arlington, Arlington, TX, USA
[2]Mathematics, University of Texas at Arlington, Arlington, TX, USA
[3]Department of Computer Science, The University of Georgia, Athens, 30602, USA
`lu.zhang2@mavs.uta.edu`

**Abstract.** For decades, a variety of predictive approaches have been proposed and evaluated in terms of their prediction capability for Alzheimer's Disease (AD) and its precursor – mild cognitive impairment (MCI). Most of them focused on prediction or identification of statistical differences among different clinical groups or phases (e.g., longitudinal studies). The continuous nature of AD development and transition states between successive AD related stages have been overlooked, especially in binary or multi-class classification. Though a few progression models of AD have been studied recently, they were mainly designed to determine and compare the order of specific biomarkers. How to effectively predict the individual patient's status within a wide spectrum of continuous AD progression has been largely overlooked. In this work, we developed a novel learning-based embedding framework to encode the intrinsic relations among AD related clinical stages by a set of meaningful embedding vectors in the latent space (*Disease2Vec*). We named this process as *disease embedding*. By disease embedding, the framework generates a *disease embedding tree (DETree)* which effectively represents different clinical stages as a tree trajectory reflecting AD progression and thus can be used to predict clinical status by projecting individuals onto this continuous trajectory. Through this model, DETree can not only perform efficient and accurate prediction for patients at any stages of AD development (across five clinical groups instead of typical two groups), but also provide richer status information by examining the projecting locations within a wide and continuous AD progression process. (Our code has been released[1].).

## 1 Introduction

Alzheimer's disease (AD) is the most common cause of dementia that cannot be prevented, cured, or even slowed. Earlier studies have shown that AD pathogenesis involve widespread alterations in brain structure and/or function, such as hippocampi [1], gray matter atrophy [2], white matter disruption [3] and abnormal functional connectivity in default mode network (DMN) [4]. Based on these brain alterations, many approaches have been developed for early diagnosis of AD and its prodromal stage – mild cognitive impairment (MCI), such as voxel-based analysis [5], tract-based spatial statistics [6], and recently developed machine learning/deep learning based models [7-9]. However, as a neurodegenerative disorder with a long pre-clinical period, the spectrum of AD spans from clinically asymptomatic to severely impaired [10]. For example, heterogeneity in clinical presentation, rate of atrophy and cognitive decline [11] may occur in the prodromal stage of AD [12]. Furthermore, individual variations may also contribute to the heterogeneity of AD: earlier studies suggested that the gap between cognitive function and brain pathology (i.e., cognitive reserve) is typically larger in highly educated individuals [13]. In general, traditional predictive approaches (e.g., classification-based models) may be limited in describing the continuum of AD development and individual variations in clinical prediction. To address this potential limitation, hypothetical models [14] for AD progression have been proposed and followed by various progression studies using cross-sectional or short-term follow-up dataset. These attempts include regression-based models [15], event-based models [16] and other computational models [17]. Nevertheless, most of them were designed to determine the order of biomarkers. Because these models consider different measures or biomarkers separately, they create a different trajectory for each biomarker. Consequently, different models and assumptions may lead to inconsistent results and interpretations [16]. More importantly, previous AD progression models are based on population analysis, they cannot be directly used for individualized diagnosis and prediction.

---

[1]We have released our code on GitHub and the code has been examined and used by other collaborators. We will add the link of the code into the paper upon acceptance.

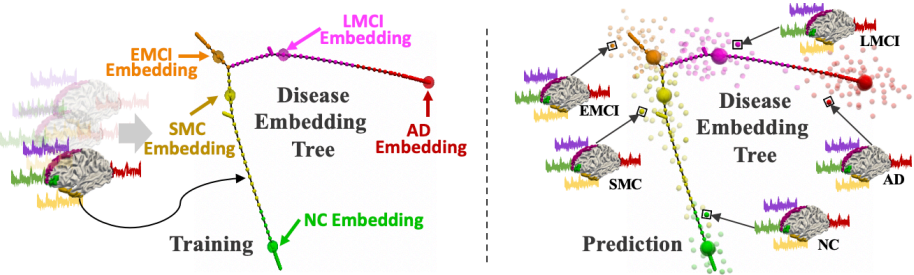

**Fig. 1. Training:** we used functional connectivity as input and learned a Disease Embedding Tree (DETree) to model the entire progression of AD in the latent space. In the tree structure, each small bubble represents a single subject, and the colors indicate different clinical groups, including normal control – NC (*green*), significant memory concern – SMC (*yellow*), early MCI – EMCI (*orange*), late MCI – LMCI (*pink*) and AD (*red*). Each edge in the DETree indicates that the connecting two nodes have higher similarity in the latent space. The five bigger bubbles represent the learned group embeddings. **Prediction:** During the prediction, new patients will be projected into the latent space which are represented as scattered bubbles. The color of the bubble indicates the true label, the location of the bubble shows its state in the entire development process from NC to AD, and the prediction of the bubble is based on the nearest embeddings.

Recent advances in deep modeling have triggered a new era in representation learning field, and a variety of powerful embedding learning algorithms have been proposed. Inspired by the remarkable success of word embedding methods in Natural Language Processing (NLP) field [18], in this work, we designed a new learning-based embedding framework to encode the intrinsic relations among AD related clinical stages by a set of meaningful embedding vectors in the latent space (*Disease2Vec*). The relations between different embedding vectors, in turn, can appropriately represent the relations between the clinical stages. Figure 1 illustrates the main idea of our framework. During the training process, the input feature of individuals will be transformed into a latent space to obtain individual embeddings. In the same latent space, a set of embeddings for different clinical groups will be parameterized and learned. Then by embedding marching, individuals will be classified to the group which the nearest group embedding belongs to. To effectively represent the continuous manner of AD development, a novel *ordered embedding constraint* was proposed to guide the embedding learning process. After the model was well trained, we obtained a tree-based trajectory that can effectively integrate the AD progression modeling and individual prediction. We named it as *Disease Embedding Tree (DETree)*. During the prediction process, by projecting individuals onto the continuous trajectory, the learned DETree can not only assign the clinical group to new patients but also show their clinical status in the entire development process from NC to AD. With the learned DETree, our model achieves a relatively high classification accuracy – 77.8% for multi-class classification (NC vs. SMC vs. EMCI vs. LMCI vs. AD), compared to other established machine learning methods and reported results in the literature [7, 13, 21, 22].

## 2 Methods

### 2.1 Data

**Data Description and Preprocessing.** We used 266 subjects (60 NC, 34 SMC, 51 EMCI, 62 LMCI and 59 AD) from the ADNI dataset (http://adni.loni.usc.edu/). Each subject has both structure MRI (T1-weighted) and resting state fMRI (rs-fMRI) data. For T1-weighted MRI, FOV = 240 × 256 × 208mm$^3$, voxel size = 1.0 mm isotropic, and TR = 2.3s. The rs-fMRI data has 197 volumes, FOV = 220 × 220 × 163mm$^3$, voxel size = 3.3mm isotropic, TR = 3s, TE = 30ms and flip angle = 90°. The first 6 volumes were discarded during preprocessing procedures to ensure magnetization equilibrium. We applied skull removal for both T1 and rs-fMRI modalities. And for rs-fMRI images, we applied spatial smoothing, slice time correction, temporal pre-whitening, global drift removal and band pass filtering (0.01-0.1 Hz). All these preprocessing steps are implemented using FMRIB Software Library (FSL) (https://fsl.fmrib.ox.ac.uk/fsl/fsl-wiki/) FEAT. For T1 images, we conducted segmentation by FreeSurfer package (https://surfer.nmr.mgh.harvard.edu/). After the segmentation, we adopted the Destrieux Atlas for ROI labeling, and the brain cortex is partitioned into 148 regions.

**Generation of Functional Connectivity.** We calculated averaged fMRI signal for each brain region. Previous studies [19] suggested that for rs-fMRI 14 time points (when TR=2s) are sufficient to capture functional dynamic patterns. To enlarge the dataset, we divided the signal into four non-overlapping segments and each segment has 45 time points. We used Pearson Correlation Coefficient to calculate functional connectivity for each of the four groups of the signal segments and obtained four functional connectivity matrices for each subject. These functional connectivity matrices were vectorized and used as input of our model.

## 2.2 Method Overview

We proposed a DETree framework to represent the continuum of AD development process as a tree structure embedded in a latent space. Here, an embedding is an abstract representation defined in latent space that is associated with a specifical clinical stage. We parameterized a group of embeddings as hidden variables in latent space (*Sec. 2.3*) and used the order information of clinical groups (NC → SMC → EMCI → LMCI → AD) to guide the embedding process (*Sec. 2.4*). In general, the proposed model aims to learn a deep representation of the input signals in a latent space that is specially optimized for both tasks simultaneously: the individual prediction and the AD progression learning. As a result, on the learned tree structure, the patients with similar clinical status are close and distant otherwise. Moreover, DETree can predict the clinical stage for a new patient by projecting it to the proper location on the learned tree structure (*Sec. 2.4*). Next, we will present the details of DETree and its predictive capability for new patients.

## 2.3 Disease Embedding Learning

Let $\{(x_i, y_i)\}_{i=1}^n$ be the training data consisting of $n$ labeled data with the $i^{th}$ input $x_i \in \mathcal{R}^d$ and class label $y_i \in \{1, \dots, C\}$ with $C$ disease stages. To maintain representative instances for different disease stages, we parameterized and learned a set of embeddings in latent space and used embedding matching for classification.

First, we adopted a non-linear function $h(x, \theta): \mathcal{R}^d \to \mathcal{R}^k$ to transform any given input $x \in \mathcal{R}^d$ to a latent space $\mathcal{R}^k$ with model parameter $\theta$. Multi-layer perceptron (MLP) network can be the potential function. And then, we defined a set of embeddings as $\mathcal{E} = \{e_{i,j} \in \mathcal{R}^k | i = 1, 2, \cdots, C; j = 1, 2, \cdots, K\}$ where $K$ is the number of the embeddings in each class. With the help of the non-linear transformation function and the set of embeddings in the latent space, we can make prediction for any given data. Specifically, given an input data $x \in \mathcal{R}^d$, we first got its representation (latent feature) $h(x, \theta)$ in latent space, then we compared the latent feature with all embeddings and classified it to the category $y$, which the nearest embedding belongs to:

$$y = \operatorname*{argmin}_{i \in \{1,2,\cdots C\}} \min_{j \in \{1,2,\cdots K\}} \|h(x, \theta) - e_{i,j}\|_2^2 \tag{1}$$

The network parameters $\theta$ and embeddings $\mathcal{E}$ can be trained jointly in an end-to-end manner, which can make the MLP model and embeddings interact with each other for better performance. To train the model, we need to define a proper loss function such that 1) it is differentiable with respect to $\theta$ and $\mathcal{E}$, and 2) it should be closely related to the classification accuracy.

**Embedding Learning Based Cross Entropy Loss.** In our DETree model, we used distance to measure the similarity between the input samples and the embeddings. The class label of the embedding $e_{i,j}$ can be denoted by $y_{i,j}$, to indicate the $j^{th}$ embedding of class $y_i$. Thus, the probability of an input $x$ belongs to class $y_i$ (i.e., $e_{i,j}$ is the nearest embedding of $x$) is formulated as:

$$P(y_{i,j}|x) = \frac{\exp\{-\alpha\|h(x,\theta)-e_{i,j}\|_2^2\}}{\sum_{l=1}^{C}\sum_{m=1}^{K}\exp\{-\alpha\|h(x,\theta)-e_{l,m}\|_2^2\}} \quad (2)$$

where $\alpha$ is a hyper-parameter that controls the hardness of distance in probability assignment. Given the definition of $P(y_{i,j}|x)$, we can further define the probability of an input $x$ belonging to the category $c \in \{1,2,\cdots,C\}$ as:

$$P(c|x) = \sum_{j=1}^{K} P(y_{c,j}|x) \quad (3)$$

Then, we defined a classification loss function based on the probability $P(c|x)$ and named it as embedding learning based cross entropy loss given by:

$$\mathcal{L}_{\mathcal{E}}((x,y);\theta;\mathcal{E}) = -\frac{1}{C}\sum_{c=1}^{C} \mathbb{I}(c=y)\log P(c|x) \quad (4)$$

where indicator function $\mathbb{I}(c=y)$ is 1 if predicator $c=y$ is true and 0 otherwise.

From (2), (3) and (4), we can see that minimizing the embedding learning based cross entropy loss essentially means decreasing the distance between the latent feature $h(x,\theta)$ of input sample $x$ and the embedding vector, which comes from the genuine category of $x$. By this way, the distance of two input samples at the same disease stage will be small in the latent space and the disease related representative embeddings can be automatically learned from data.

To improve the generalization performance and prevent over-fitting, we also proposed a new embedding-based regularization term:

$$\mathcal{L}_{\mathcal{ER}}((x,y);\theta;\mathcal{E}) = \|h(x,\theta) - e_{y,*}\|_2^2 \quad (5)$$

where $e_{y,*}$ is the closest embedding of $h(x,\theta)$ with class label $y$. The regularization term pulls the latent feature $h(x,\theta)$ of input sample $x$ close to their corresponding embedding, making the latent features within the same class more compact, so it is beneficial for classification.

### 2.4 Ordered Embedding Constraint

The class labels $y$ provides not only the separability of their inputs, but also the underlying relationship of the clinical groups, which corresponds to different disease stages during the progression of AD. It is generally assumed that the ordering of the clinical groups is NC → SMC → EMCI → LMCI → AD. Even though the ordering of each input sample is unknown, the ordering of the classes can still provide valuable information to guide the embedding learning. To take advantage of this prior knowledge, we constructed an affinity matrix $\mathcal{A} = \left[a_{(i,j),(i',j')}\right] \in \mathcal{R}^{N \times N}$ for the similarity among embedding class labels. $N = C \times K$ is the total number of embeddings. $a_{(i,j),(i',j')} = 1$ if the $(i,j)^{th}$ embedding and the $(i',j')^{th}$ embedding are from the same class, that is $y_i = y_{i'}$, $a_{(i,j),(i',j')} = 0.5$ if $y_i$ is the neighbor of $y_{i'}$ in the ordering of class labels, and 0 otherwise.

To leverage this prior information for learning the path of AD progression, we added an additional neural network layer with softmax function onto the embeddings to link the latent feature of embeddings and the clinical groups. As a result, the output probability of embedding $e_{i,j}$ belonging to the class $c$ is formulated as:

$$O_c(e_{i,j}; W, b) = \frac{\exp\{(W_c^T e_{i,j} + b_c)\}}{\sum_{l=1}^{C} \exp\{(W_l^T e_{i,j} + b_l)\}} \quad (6)$$

where $\{W_l, b_l\}$ are the parameters of the neural network layer. The final prediction is:

$$y_{i,j} = \underset{c \in \{1,2,\cdots C\}}{\mathrm{argmax}}\, O_c(e_{i,j}; W, b) \quad (7)$$

According to (6) and (7), the classification loss of embeddings is defined as:

$$\mathcal{L}_O(e_{i,j}; W, b) = -\frac{1}{C}\sum_{c=1}^{C} \mathbb{I}(c = i)\, log\, O_c(e_{i,j}; W, b) \quad (8)$$

Then, we proposed the following regularization term to incorporate the ordering information of the class labels in terms of the affinity matrix $\mathcal{A}$ based on the manifold assumption: if two labels are similar, their probabilities of predictions should be close. The regularization term is then formulated as:

$$\mathcal{L}_{OR}(\mathcal{E}; W, b) = trace(OL_y O^T) \quad (9)$$

where $O = [O_1; O_2; \cdots; O_C] \in \mathcal{R}^{C \times (C \times K)}$ and $O_c = \left[O_c(e_{i,j}; W, b)\right]_{\{i,j\}} \in \mathcal{R}^{1 \times (C \times K)}$, $\forall c$, $L_y = \mathcal{D} - \mathcal{A}$ is the graph Laplacian matrix of $\mathcal{A}$ and $\mathcal{D}$ is the degree matrix of $\mathcal{A}$.

Together with (4), (5), (8), (9) in hand, we are now ready to formulate our DETree model with the loss function defined as:

$$\mathcal{L} = \sum_{i=1}^{n}\left[\mathcal{L}_{\mathcal{E}}((x_i, y_i); \theta; \mathcal{E}) + \beta \mathcal{L}_{\mathcal{ER}}((x_i, y_i); \theta; \mathcal{E})\right]$$
$$+ \gamma \sum_{i=1}^{C}\sum_{j=1}^{K} \mathcal{L}_O(e_{i,j}; W, b) + \delta \mathcal{L}_{OR}(\mathcal{E}; W, b) \quad (10)$$

This loss function (10) is derivable with respect to $\theta$, $\mathcal{E}$, $W$ and $b$. The whole model can be trained in an end-to-end manner. After the model was trained, we used Kruskal's algorithm to create a minimum spanning tree over latent features $h(x, \theta)$. This tree was learned by disease embedding learning, and we named it as *Disease Embedding Tree*.

For a new patient $x$, DETree can provide two sets of predictions. First, the probability of assigning $x$ to each of the given clinical group can be obtained via (3), and the best prediction can be made according to (1). The second prediction is to decide the location of the patient on the learned DETree by $h(x, \theta)$, which reflects the stage where the patient is in the whole progression of the disease.

## 3 Results

### 3.1 Experimental Setting

**Data Setting.** In this work, we used 266 subjects (60 NC, 34 SMC, 51 EMCI, 62 LMCI, 59 AD) to do experiment. Based on Sec. 2.1 each subject has four functional matrices and we obtained 1064 data samples in total. In our experiments, the training, validation, and testing datasets were split according to subjects, that is, four matrices of the same subject will be divided into the same dataset. As the functional matrix is symmetric, to reduce the redundant data, we used the vectorized upper triangle of each matrix as input features.

**Model Setting.** In this work, the non-linear function $h(x, \theta)$ was implemented by six-layer MLP. The dimensions of the MLP are 1024-512-256-64-16-$k$, where $k$ is the dimension of the latent space (*Sec. 2.3*). We tested $k$ = 5, 10, 15, 20, 25. We showed the results of $k$ = 25 which gives the best classification performance in Sec. 3.2 and Sec. 3.3, and compared the results of $k$ = 5, 10, 15, 20 and 25 in Sec. 3.4. C = 5 is the number of classes (NC/SMC/EMCI/LMCI/AD). Activation function Relu and batchnorm were used at each layer. The other four hyper-parameters are $\alpha$ = 1.0, $\beta$ = 0.001, $\gamma$ = 1.0 and $\delta$ = 1.0. The number of the embeddings in each class is $K$ = 1. The entire model was trained in an end-to-end manner. During the training process, the Adam optimizer was used to train the whole model with standard learning rate 0.001, weight decay 0.01, and momentum rates (0.9, 0.999).

**Table 1.** Classification Performance of DETree and Four Other Machine Learning Methods. The red color and blue color represent the best and the second-best results, respectively.

| Method | F1 | | | | | | Acc (All) |
|---|---|---|---|---|---|---|---|
| | All | AD | LMCI | EMCI | SMC | CN | |
| SVM | 0.658 ±0.05 | 0.684 ±0.06 | 0.641 ±0.05 | 0.630 ±0.05 | 0.659 ±0.05 | 0.675 ±0.04 | 0.672 ±0.03 |
| KNN | 0.526 ±0.05 | 0.427 ±0.09 | 0.576 ±0.05 | 0.546 ±0.07 | 0.547 ±0.08 | 0.535 ±0.04 | 0.532 ±0.04 |
| Logistic Regression | 0.642 ±0.03 | 0.651 ±0.06 | 0.634 ±0.04 | 0.627 ±0.05 | 0.649 ±0.05 | 0.651 ±0.03 | 0.649 ±0.02 |
| Random Forest | 0.422 ±0.03 | 0.377 ±0.05 | 0.516 ±0.07 | 0.426 ±0.04 | 0.436 ±0.06 | 0.354 ±0.04 | 0.428 ±0.03 |
| DETree (Our) | **0.777 ±0.01** | **0.785 ±0.03** | **0.762 ±0.03** | **0.801 ±0.06** | **0.765 ±0.02** | **0.773 ±0.03** | **0.778 ±0.02** |

### 3.2 Classification Performance

In this section, we showed the classification performance of the proposed DETree. For fair comparisons, we used two strategies to compare the proposed method with other widely used methods. First, we repeated experiments 5 times with random seeds to compare the results with other four broadly used machine learning methods including support vector machine (SVM), k-nearest neighbors (KNN), logistic regression and random forest. The classification performance was measured by $F_1$ scores: $F_1 = 2 \times \frac{precision \times recall}{precision + recall}$ and accuracy (Acc). The results are showed in Table 1. We can see that the $F_1$ scores of DETree model is over 0.75 which is more than 10% higher than the second-best results (highlighted by blue). And for some classes it can reach 0.80, which is outstanding in multi-class classification of AD and significantly outperforms the other four methods. Second, we compared the multi-class classification performance with five latest deep learning methods on AD and reported the results in Table 2. As shown in Table 2, [20] obtains a very high $F_1$ score for AD group, however the $F_1$ scores for other groups are much lower. The total accuracy of [13] and [7] are very close to our results, however, they only included three classes while we considered five classes in this work.

**Table 2.** Classification Performance of DETree and Other Deep Learning Methods. cMCI/pMCI: MCI patients who converted to AD within 36 months; sMCI: MCI patients who didn't convert to AD within 36 months. EMCI/LMCI: early/late mild cognitive. The red color and blue color represent the best and the second-best results, respectively.

| Work | Modality | Subjects | Method | Performance |
|---|---|---|---|---|
| Amorosoa et al. (2018) [20] | Predefined Features | 60AD, 60HC, 60cMCI, 60MCI | Deep Random Forest | $AD: F_1 = 0.805$<br>$sMCI: F_1 = 0.305$<br>$cMCI: F_1 = 0.518$<br>$NC: F_1 = 0.525$ |
| Zhou et al. (2018) [21] | MRI, PET, SNP | 190AD, 226HC, 157pMCI, 205sMCI | Multi-modal Fusion | $AD: Acc = 0.574$<br>$pMCI: Acc = 0.622$<br>$sMCI: Acc = 0.342$<br>$NC: Acc = 0.625$ |
| Brand et al. (2019) [22] | MRI, SNP | 412 (AD/MCI/NC) | Joint Regression-Classification | $AD: F_1 = 0.566$<br>$MCI: F_1 = 0.513$<br>$NC: F_1 = 0.683$ |
| Lei et al. (2020) [13] | MRI | 192AD, 402MCI, 220NC | Multiple templates, Adaptive Feature Selection | $Total: Acc = 0.775$<br>(NC/MCI/AD) |
| Wang et al. (2020) [7] | rs-fMRI | 253CN, 45EMCI, 88LMCI | Deep Autoencoder | $Total: Acc = 0.73$<br>(NC/EMCI/LMCI) |
| DETree (Our) | rs-fMRI | 59AD, 62LMCI, 51EMCL, 34SMC, 60NC | DETree | $AD: F_1 = 0.785$<br>$LMCI: F_1 = 0.762$<br>$EMCI: F_1 = 0.801$<br>$SMC: F_1 = 0.765$<br>$NC: F_1 = 0.773$<br>$Total: Acc = 0.778$ |

### 3.3 The learned Embedding Tree

An important contribution of our DETree is using the learned tree structure to represent the entire spectrum of AD progression. As mentioned in the model setting section, the learned DETree is in a high dimensional latent space ($k$ = 5,10,15,20,25). To simultaneously visualize the learned tree structure and maintain the intrinsic relations among different clinical groups, we used the adjacent matrix weighted by distance to draw a connection DETree. We used NetworkX Software (https://networkx.org/) to display the tree structure: it focuses on the connection relationships and the relative distance of the vertices but ignores their actual coordinates in the original latent space. We showed the connection DETree in Fig. 2. The first row of Fig. 2 is the learned five connection DETree and each one is from one result of the five runs of the experiments. In the tree structure, each small bubble represents a single subject, and the colors indicate different AD related groups. Each edge in the connection DETree represents the connecting two nodes have higher similarity in the latent space. The five bigger bubbles represent the learned embeddings, and the color indicates the class they belong to. The learned DETree structure displays a trajectory of AD progression: it starts from NC, goes through SMC, EMCI, LMCI and eventually ends with AD. In the second row, to visualize the prediction results we projected the new patients onto the connection DETree with the following projection steps: 1) using the well trained MLP model to project all the new samples to the latent space to obtain the latent features ($h(x, \theta)$); 2) assigning the predicted group to each new sample with formula (1); 3) projecting each new sample within a neighborhood of the corresponding embedding. The location is randomly assigned but the relative distance to other embeddings and samples are maintained.

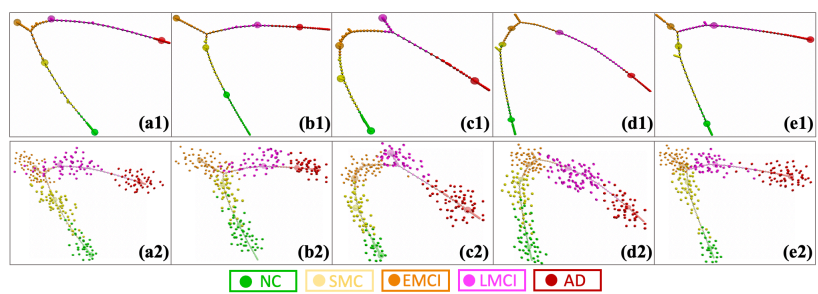

**Fig. 2.** First Row: visualization of connection DETree from multiple clinical groups including NC, SMC, EMCI, LMCI and AD. Second Row: visualization of the prediction of new patients by projecting on the connection DETree.

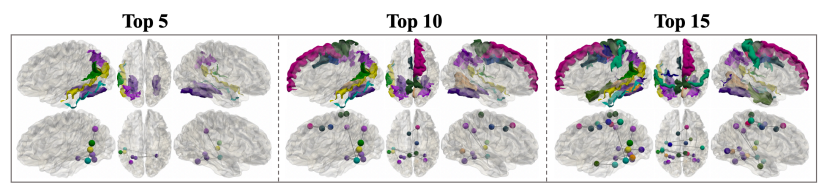

**Fig. 3.** Top connectivity that contributes most to the learned DETree structure. In each subfigure, the top and bottom rows display the involved brain regions and connectivity, respectively.

In this work, we used functional connectivity to learn the DETree. To further explore which functional connectivity contributes most to the learned tree structure, we ordered them with Laplacian score [23] during the learning of DETree structure. Figure 3 shows the top 5, 10 and 15 connectivity that have the most contributions. In each subfigure, the first row shows the brain regions involved in the connectivity. The second row shows the connectivity, and the corresponding regions are represented by bubbles with the same color. Most of the regions in Fig. 3 are reported in previous studies for the close relationship to AD, such as the regions in frontal lobe and temporal lobe [7-9].

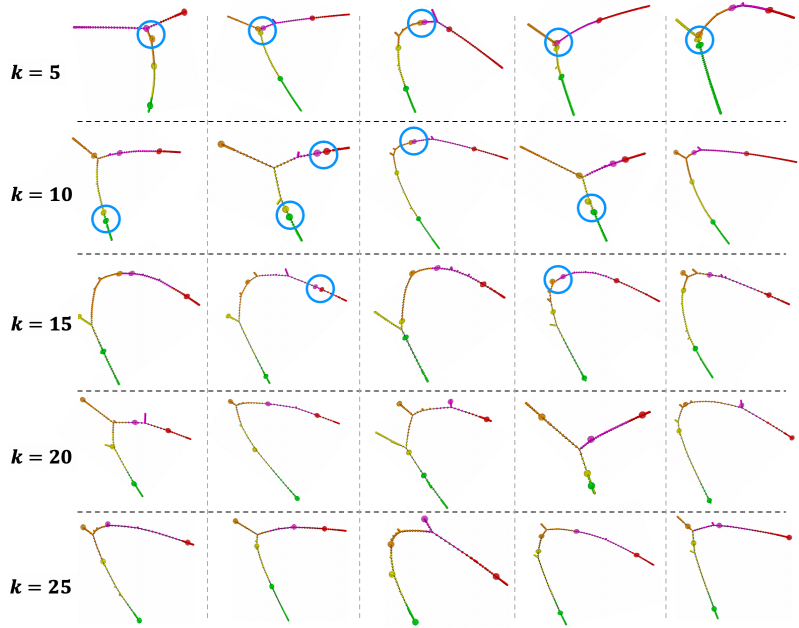

**Fig. 4.** Connection DETree structures correspond to different dimensions of the latent space. The blue circles are used to highlight the embeddings from different clinical groups with small distance.

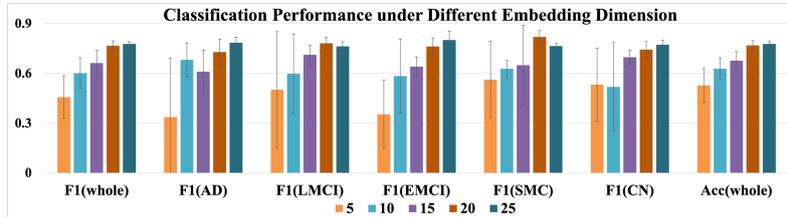

**Fig. 5.** Classification performance with different dimension of the latent space.

### 3.4 Reproducibility of Deep Embedding Tree Learning

In our DETree model, the hyper-parameter that have the most important in- fluence on the DETree structure is $k$, the dimension of latent space. We tested $k$ = 5,10,15,20 and 25 and reported the classification performance as well as the connection DETree structure with different $k$ in Fig. 5 and Fig. 4, respectively. Figure 5 shows that the classification performance improves when increasing $k$ and reaches the peak at some values between 20 and 25. From Fig. 4 we can see that if $k$ is too small (corresponding to lower dimensional latent space), the distances among different embeddings incline to be small. Insufficient dissimilarity between embeddings may limit the capability for DETree when representing multiple clinical stages in AD progression and compromise the prediction performance in estimating new samples. We used blue circle to highlight different embeddings with small distance in Fig. 4.

## 4 Conclusion and Discussion

Here we proposed a novel DETree framework to represent the continuum of AD development. The learned DETree structure displays a trajectory of AD progression and achieves a high prediction performance over 77.8% for multiple AD related stages. The learned DETree can not only predict the clinical status of individual patient, but also provide more information of the patient's state within the entire spectrum of AD progression. We summarize the advantages and limitations of current work as following:

**DETree is a general framework for modeling continuous diseases development.** In this work, we only applied DETree to Alzheimer's disease, but it is a general framework that can be applied to a wide range of disease related studies. Any disease that presents multiple clinical stages in the development can use the DETree framework by feeding suitable features into the model and adjusting the MLP structure according to the input features and tasks. More important, by modifying the affinity matrix A, the prior knowledge about the disease can be easily introduced into the DETree model.

**DETree is not limited to the classification task.** By adjusting the additional neural network layer in Sec. 2.4, the DETree can be applied to regression problems. For example, by replacing the discrete clinical labels with continuous clinical scores (i.e., Mini-Mental State Exam (MMSE) score), the proposed classification framework can be converted to a regression-based model.

**The number of embeddings in each class is flexible.** In this work we set the number of embeddings for each clinical group to be 1 ($K$=1) based on our application. However, through modifying the value of $K$, the proposed DETree can be extended by allowing multiple embeddings in each group. For instance, we can introduce multiple embeddings to represent subtypes [11] of AD and each subtype may associated with different clinical presentations (features).